\documentclass[cameraready]{vgtc}                      

\graphicspath{{figures/}{pictures/}{images/}{./}} 

\usepackage{times}                     

\usepackage{tabu}                      
\usepackage{booktabs}                  
\usepackage{lipsum}                    
\usepackage{mwe}                       

\usepackage{mathptmx}                  

\onlineid{0}

\vgtccategory{Research}

\vgtcinsertpkg

\title{\textsc{GenLARP}: Enabling Immersive Live Action Role-Play through LLM-Generated Worlds and Characters}

\author{
    Yichen Yu$^{1*}$\thanks{Equal Contribution}\thanks{Emails: yyu55@ncsu.edu} \\
    \scriptsize North Carolina State University \\
    \scriptsize Georgia Institute of Technology
    \and
    Yifan Jiang$^{2*}$\thanks{Emails: yjiang54@u.rochester.edu} \\
    \scriptsize University of Rochester
    \and
    Mandy Lui$^{3}$\thanks{Emails: mlui@u.rochester.edu} \\
    \scriptsize University of Rochester
    \and
    Qiao Jin$^{4}$ \thanks{Emails: qjin4@ncsu.edu}\\
    \scriptsize North Carolina State University
}

\teaser{
  \centering
  \includegraphics[width=\textwidth]{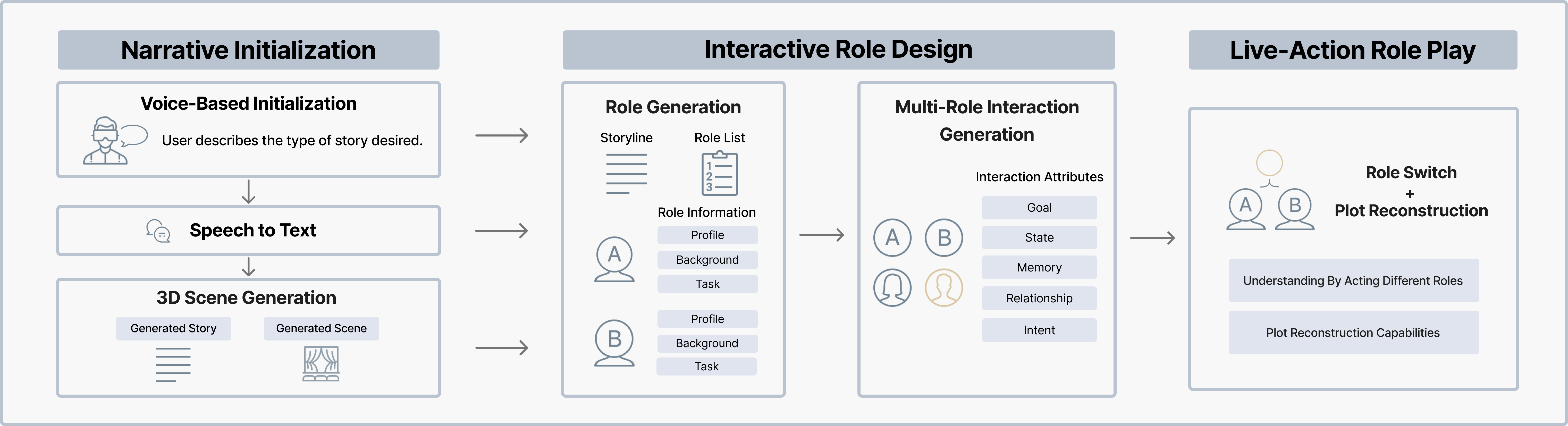}
  \caption{The system pipeline of \textsc{GenLARP}, from narrative input to immersive VR-based live action role-play (LARP).}
  \label{fig:teaser}
}

\abstract{
     We introduce \textsc{GenLARP}, a virtual reality (VR) system that transforms personalized stories into immersive live action role-playing (LARP) experiences. \textsc{GenLARP} enables users to act as both creators and players, allowing them to design characters based on their descriptions and live in the story world. Generative AI and agents powered by Large Language Models (LLMs) enrich these experiences.
} 

\keywords{Virtual Reality, LARP, Human-Computer Interaction, Agent, AI-Driven NPCs, Narrative Generation, Immersive.}

\begin{document}


\firstsection{Introduction}

\maketitle

Live-action role-playing games (LARPs) are a form of participatory storytelling in which individuals take on fictional roles, both physically and emotionally, and engage in a structured narrative. In contrast to traditional media where the audience is merely passive observers, live action role-playing games transform participants into co-creators of the experience. They are not only witnesses to the story, but also experiencers of it. This shift from observation to hands-on participation promotes a deeper emotional engagement with the culture, history, or fantasy world.

LARP is a great form of experience because it offers a unique and powerful mode of interaction. First, it allows participants to become deeply immersed, to use their imagination, and to shape the unfolding events through their own choices ~\cite{Pothong2021,Eddy2020}. Second, by taking on roles different from their own, participants can explore different perspectives and express potential aspects of their own identity~\cite{Pothong2021}. Third, it functions as a social simulation: through collaborative improvisation, participants engage with issues of power, morality, and interdependence in ways that reflect the intricacies of the real world~\cite{Pothong2021}. When based on classic stories or cultural texts, LARP allows participants to reinterpret familiar narratives from new perspectives.

However, traditional LARP poses an obstacle to wider popularization. It relies heavily on the real-time coordination of multiple participants and often requires elaborate material preparation, including elements such as costumes, props, physical sets, and facilitation ~\cite{Tychsen2006,Lacanienta2022}. These factors limit the ease and feasibility of LARP in everyday use.

To address these challenges, we propose \textsc{GenLARP}, a system that utilizes generative AI and VR to support immersive reenactments in virtual environments. Instead of requiring a large group of participants or expensive physical infrastructure, \textsc{GenLARP} enables learners to enter virtual worlds based on story scenarios they describe. In these environments, users take on roles, interact with large-scale language model (LLM)-driven agents, and experience information about different characters' states, relationships, intent, etc., through LLM-driven dialog and interaction.

\section{System}
The \textsc{GenLARP} system consists of three main modules that are responsible for transforming user input into a structured narrative, establishing interaction mechanisms, and enabling immersive experiences.

\subsection{Narrative Initialization}
This module is the entrance of the whole system, which is responsible for receiving the natural language text input by the user and transforming it into semantic information with narrative structure through language modeling. A unified semantic extraction process is designed to recognize key elements in the input text and generate textual information including spatial context, temporal cues, sources of conflict, role distribution, task motivation, and so on. The intermediate representations are further formatted as structured prompts, which are used to drive the subsequent story generation, characterization and 3D scene construction~\cite{GenEx2025}. The structured prompt follows a unified data protocol, which enables semantically consistent and uniformly formatted information transfer between sub-modules.

In terms of design, the module takes into account the non-standard and open nature of the LARP creation process, allowing users to freely express story ideas without having to follow a fixed template~\cite{Bowman2014}. For inputs that are vague, incomplete or missing important elements, the system automatically supplements the missing parts such as character structure, conflict design, spatial layout, etc. through mechanisms such as contextual reasoning, story template completion, and typical narrative pattern fitting, so as to ensure that the subsequent generation of the link has a complete input semantics. This module not only assumes the function of story engine, but also provides the whole system with unified narrative semantic boundaries and generative tone, which ensures that the LARP experience has contextual consistency and dramatic tension.

\subsection{Interactive Role Design}
Each role is driven by a large language model with independent internal state management mechanisms, including memories, affective variables, motivations and beliefs~\cite{ran2025bookworldnovelsinteractiveagent}. Characters' behavioral responses are not only dependent on the current context, but also on their individual perceptions, social locations, and goal tendencies, resulting in interaction patterns with individualized logic.

The characters are built in a plot-driven social structure, and there may be multi-level interaction mechanisms such as power relations, task dependency, trust thresholds, misunderstanding and competition~\cite{Tychsen2006}. The system supports users to freely switch between multiple characters or maintain a single character perspective, enter into the internal character experience, and engage in context-based dialogues, transactions, cooperation, conflicts and other behaviors with other AI characters, reconstructing the improvisation atmosphere under real social dynamics.

The module provides a complete state tracking, behavior path recording and belief dynamic updating mechanism to ensure that the character maintains a reasonable and consistent behavior logic in multiple rounds of interactions, avoiding uncontrolled generation or semantic breaks. The relationship structure between the characters can also evolve dynamically during the runtime, for example, due to specific events leading to the disintegration of alliances, the breakdown of trust, or the reconstruction of common goals, thus constituting an ever-changing space that fully reflects the logic of generating roles prior to the script.

\subsection{Live-Action Role Play}
The module presents the aforementioned generated content in front of the user in an immersive way, supporting the user to move, observe, and interact freely in the scene in the first-person mode through the VR device, and triggering real-time feedback with the character, objects, and environment in the space~\cite{Barab2005}. Based on the user's position in the virtual environment, behavior path, interaction history, perspective preference and other information, the system adjusts the scene state and character behavior response in real time, and builds a dynamic interaction feedback mechanism.

The system allows users to switch roles and re-experience the same storyline from different positions and perspectives, so as to perceive the latent multi-dimensional conflicts and asymmetric knowledge structure among the characters in the narrative. This mechanism reinforces the experience logic of contextual reflection.

In order to support non-linear narratives, the module provides localized plot regression and branch reconstruction capabilities, allowing users to revisit key event nodes or explore untriggered character interactions and side quests without interrupting the overall story flow~\cite{Falk2004}. The system evaluates the plot heat and interaction density in real time according to user behavior, and automatically optimizes the rhythm of character appearances, the density of scene interactions, and the rhythm of narrative advancement to adapt to the user's behavioral patterns and exploration intentions.

Ultimately, the module serves as the interaction hub between the user and the system, integrating the logical consistency between semantic inputs, character behaviors and spatial narratives to ensure that the entire LARP experience is not only generative and interactive, but also immersive and experience-able.

\section{Implementation}
\textsc{GenLARP} is implemented as a modular system that combines LLM-based semantic processing with real-time 3D rendering in Unity. Narrative input is processed using GPT-4o to extract structured hints, which are then fed into the SynCity~\cite{engstler2025syncity} framework for generating 3D scenes. SynCity supports tile-based, hint-conditional scene construction for coherent, scalable virtual environments. The generated asset is imported in real-time into Unity, where a character agent driven by a GPT-based API interacts with the user in real-time.

\section{Future Work}
While \textsc{GenLARP} demonstrates the feasibility of using generative AI and VR to support immersive narrative experiences, there are several directions for future development. First, the current system operates primarily in a single-user environment. Future iterations may explore multi-user co-creation and synchronized role-playing, enabling collaborative narrative and negotiation between participants. Second, while language modeling allows for dynamic character interactions, maintaining long-term coherence over long periods of narrative remains a challenge. Incorporating memory persistence, user modeling, and narrative planning frameworks can enhance coherence and continuity. Third, the current scene generation process is guided only by structured textual input. Integrating multimodal conditions such as sketches, soundscapes, or voice descriptions could support richer and more diverse scene construction. Finally, user experience evaluations in real-world environments are needed to better understand the impact of interactive AI-driven role-playing on imagination, identity exploration, and narrative engagement.



\bibliographystyle{abbrv-doi}

\bibliography{template}
\end{document}